\documentclass{ws-procs9x6}

\begin{document}

\title{Low-momentum nucleon-nucleon potential \\
and Hartree-Fock Calculations}

\author{L. Coraggio, A. Covello, A. Gargano, N. Itaco}
\address{Dipartimento di Scienze Fisiche, Universit\`a 
di Napoli Federico II, \\ and Istituto Nazionale di Fisica Nucleare, \\
Complesso Universitario di Monte  S. Angelo, Via Cintia - I-80126 Napoli, 
Italy}
\author{T. T. S. Kuo}
\address{Department of Physics, SUNY, Stony Brook, New York 11794}

\maketitle

\abstracts{
A low-momentum nucleon-nucleon ($NN$) potential $V_{low-k}$ is derived from
modern realistic $NN$ potentials by integrating out the high-momentum 
modes.
The smooth $V_{low-k}$ may be used as input for nuclear structure calculations
instead of the usual Brueckner $G$ matrix.
Such an approach eliminates the energy dependence one finds in the
$G$-matrix approach, allowing this interaction to be used directly in
Hartree-Fock calculations.
Bulk properties of $^{16}$O have been calculated starting from
different $NN$ potentials.
Our results, obtained including up to second order contributions in
the Goldstone expansion, are in good agreement with experiment.}

\section{Introduction}
A central problem in nuclear theory has long been the calculation of
nuclear structure properties starting from realistic nucleon-nucleon
($NN$) potentials.
There are various models for these potentials, such as the 
Bonn\cite{mach87} and CD-Bonn,\cite{cdbonn} Paris,\cite{lacombe80}
Nijmegen,\cite{stoks94} and the new Idaho potential.\cite{mach02}
They all describe the observed deuteron and $NN$ scattering data very
accurately.
However, owing to their strong repulsive core, none of them can 
be used directly in nuclear structure.
To overcome this difficulty, the Brueckner $G$ matrix has traditionally
been the starting point, but, as is well known, its energy
dependence is an undesirable feature, in particular when dealing 
with Hartree-Fock calculations.

In this paper, we make use of a different approach to renormalize the
short-range repulsion of realistic $NN$ potentials, which is motivated 
by the recent applications of effective field theory (EFT) and  
renormalization group (RG) to low-energy nuclear 
systems.\cite{lepage,kaplan98,epel98,epel99,EFT}

A fundamental theme of the RG-EFT approach is that the details of the 
short-distance dynamics are irrelevant for physics in the infrared
region.
One can therefore have infinite theories that differ substantially at
short distance, but still give the same low-energy 
physics.\cite{lepage,EFT}
In nuclear physics, various meson models for $V_{NN}$, sharing the
same one-pion tail, give the same phase shifts and deuteron binding 
energy even if they differ significantly in the treatment of 
the shorter distance pieces.

Motivated by these considerations, we derive a low-momentum $NN$
potential $V_{low-k}$\cite{bogner01,bogner02} by integrating out 
the high-momentum components of $V_{NN}$ in the sense of the 
RG,\cite{lepage,EFT} and use this smooth potential in nuclear 
structure.

In this paper, we calculate the bulk properties of $^{16}$O in the
framework of the Hartree-Fock theory, using $V_{low-k}$'s derived from
different $NN$ potentials.
The paper is organized as follows. In section 2 we describe our
method for carrying out the high-momentum integration.
Section 3 is devoted to the comparison of our results with experiment.
A summary of our conclusions is given in the last section.

\section{Outline of Calculations}
The first step in our approach is to integrate out the high-momentum
components of $V_{NN}$.
According to the general definition of a renormalization group
transformation, the decimation must be such that low-energy observables
calculated in the full theory are preserved exactly by the effective theory.
Once the relevant low-energy modes are identified, all remaining modes or
states have to be integrated out.

For the nucleon-nucleon problem in vacuum, we require that the deuteron
binding energy, the low-energy phase shifts, and the low-momentum
half-on-shell $T$ matrix calculated from $V_{NN}$ must be reproduced
by $V_{low-k}$.

The full-space nuclear Schr\"odinger equation may be written as
\begin{equation}
H \Psi_{\mu} = E_{\mu} \Psi_{\mu};~ H=H_0 + V_{NN},
\end{equation}
where $H_0$ is the unperturbed Hamiltonian, in this case the kinetic energy.
The above equation can be reduced to a model-space one of the form
\begin{equation}
PH_{\rm eff}P \Psi_{\mu} = E_{\mu} P \Psi_{\mu};~ H_{\rm eff}=H_0 + 
V_{\rm eff},
\end{equation}
where $P$ denotes the model-space, which is defined by momentum 
$k\leq k_{cut}=\Lambda$, $k$ being the relative momentum and $k_{cut}$
a cut-off momentum. 

The half-on-shell $T$ matrix of $V_{NN}$ is defined as 

\[  T(k',k,k^2) = V_{NN}(k',k) + ~~~~~~~~~~~~~~~~~~~~~~~~~~~~~~~~~~~~~~~~~~~~~~~~~~~\]
\begin{equation}
~~~~~~ + \int _0 ^{\infty} q^2 dq  V_{NN}(k',q)
 \frac{1}{k^2-q^2 +i0^+ } T(q,k,k^2 ) ~~,
\end{equation}

\noindent
and the corresponding model-space $T$ matrix given by $V_{low-k}$ is

\[  T_{low-k }(p',p,p^2) = V_{low-k }(p',p) + ~~~~~~~~~~~~~~~~~~~~~~~~~~~~~~~~~~~~~~~~~~\]
\begin{equation}
~~~~~~+\int _0 ^{\Lambda} q^2 dq  V_{low-k }(p',q)
 \frac{1}{p^2-q^2 +i0^+ } T_{low-k} (q,p,p^2)~~.
\end{equation}

\noindent
Note that for $T_{low-k }$ the intermediate states are integrated up
to $\Lambda$.

It is required that, for $p$ and $p'$ both belonging to $P$ ($p,p' \leq
\Lambda$), $T(p',p,p^2)= T_{low-k }(p',p,p^2)$.
In Refs. 11,12 it has been shown that the above requirements are
satisfied when $V_{low-k}$ is given by the folded-diagram series

\begin{equation}
V_{low-k} = \hat{Q} - \hat{Q'} \int \hat{Q} + \hat{Q'} \int \hat{Q}
\int \hat{Q} - \hat{Q'} \int \hat{Q} \int \hat{Q} \int \hat{Q} + ...~~,
\end{equation}

\noindent
where $\hat{Q}$ is an irreducible vertex function, in the sense that
its intermediate states must be outside the model space $P$.
The integral sign represents a generalized folding operation, \cite{krenc80} 
and $\hat{Q'}$ is obtained from $\hat{Q}$ by removing terms of first 
order in the interaction $V_{NN}$.

The above $V_{low-k}$ can be calculated by means of iterative techniques.
In Ref. 14 some iteration methods have been proposed which are suitable for  
non-degenerate model spaces. 
The method that we use here, which we refer to as Andreozzi-Lee-Suzuki
(ALS) method, consists in constructing solutions of the Lee-Suzuki type. 
The ALS method converges to the lowest (in energy) $d$ states of $H$, $d$ 
being the dimension of the $P$ space.
We have found this method to be very convenient for our calculations.

Since the $V_{low-k}$ obtained by this technique is non-hermitian, we
have hermitized it by using the simple and numerically convenient
procedure suggested in Ref. 14.

We have verified that the deuteron binding energy and the 
phase shifts up to the cut-off momentum $\Lambda$ are preserved by 
$V_{low-k}$.
By way of illustration, we compare in Fig. 1 some $^1 S_0$ phase shifts 
calculated using a cut-off momentum $\Lambda =2~{\rm fm} ^{-1}$ (squares), 
with those obtained using the full Idaho potential\cite{mach02} (continuous
line).
It can be seen that the phase shifts from the full $V_{NN}$ are well 
reproduced by those obtained from $V_{low-k}$ up to the value of
$E_{lab}$ corresponding to the cut-off momentum.

\begin{figure}[ht]
\centerline{\epsfxsize=3.9in\epsfbox{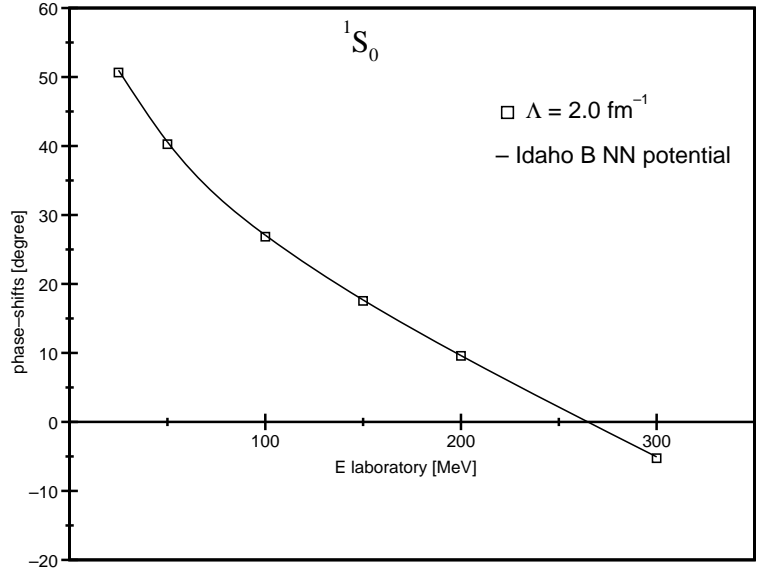}}   
\caption{Comparison of $^1 S_0$ phase shifts given by $V_{NN}$ and 
$V_{low-k}$.}
\end{figure}

This evidences the physical equivalence of $V_{low-k}$ and $V_{NN}$ in the
sense of RG.

An important issue is what value one should use for $\Lambda$.
Guided by general EFT arguments, the minimum value for $\Lambda$ must be large
enough so that $V_{low-k}$ explicitly contains the necessary degrees of freedom
for the physical system under investigation.
For nuclear structure calculations $\Lambda$ turns out to be around
$2~{\rm fm}^{-1}$.
This is consistent with the following considerations.
Most $NN$ potentials are constructed to fit empirical phase shifts up to
$E_{lab}\approx 350$ MeV.
Since $E_{lab} \le 2\hbar^2\Lambda^2/M$, $M$ being the nucleon mass,
and one requires $V_{low-k}$ to reproduce the empirical phase shifts,
a choice of $\Lambda$ in the vicinity of $2~{\rm fm}^{-1}$ seems to be
appropriate.
In our calculations we have used $\Lambda=2.1~{\rm fm}^{-1}$.

\section{Realistic Hartree-Fock Calculations}
As stated before, $V_{low-k}$ is energy independent and well-behaved,
so that one can use it directly to compute the bulk properties of
$^{16}$O in the framework of the self-consistent Hartree-Fock (HF) 
theory.
The HF equations are solved using a basis of
harmonic-oscillator wave-functions. 
\begin{figure}[ht]
\centerline{\epsfxsize=3.9in\epsfbox{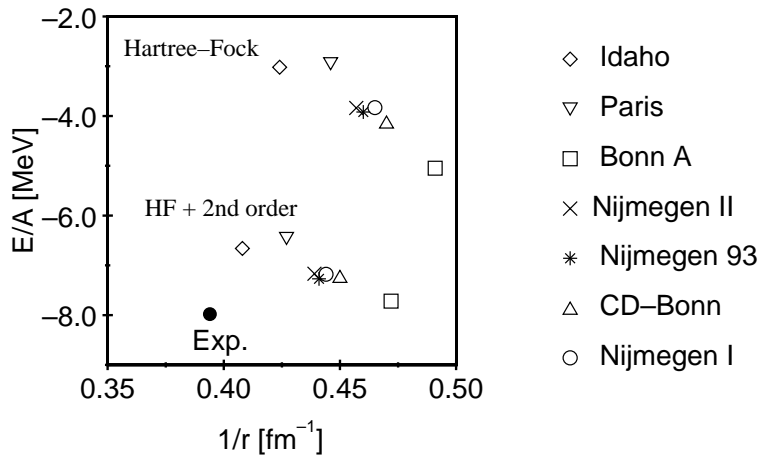}}   
\caption{Binding energy per nucleon versus inverse r.m.s. mass radius 
for $^{16}$O. The experimental data are from Refs. 15,16.}
\end{figure}
The single-particle states $ |i \rangle $ are expanded in a finite
series of oscillator wave-functions 
$ | \alpha \rangle $:
\begin{equation}
|i \rangle = \sum_{\alpha} C^i_{\alpha} | \alpha \rangle~~. 
\end{equation}
The expansion coefficients $C^i_{\alpha}$ and the oscillator parameter
$\hbar \omega$ are determined solving self-consistently the
HF equations
\begin{equation}
\sum_{\beta} \left[ \langle \alpha | T | \beta \rangle + \sum_{j_h}
\langle \alpha,j_h|V_{low-k} | \beta,j_h \rangle \right] C^i_{\beta} =
\epsilon_i C^i_{\alpha}~~.
\end{equation}
Since we assume that $^{16}$O is a spherical nucleus, the $ |i
\rangle $'s have good orbital and total angular momentum. 
The Coulomb correction is included exactly by taking the total
two-body interaction as $V_{low-k}$ plus the Coulomb potential.

Once Eqs.(11) have been solved, we use the HF wave-functions to 
calculate both the binding energy and r.m.s. mass radius.
We also evaluate the second-order contribution to the above
quantities within the framework of the Goldstone linked cluster
perturbation expansion\cite{goldstone57}.
In Fig. 2 we report the results obtained using all the $NN$
potentials mentioned in the Introduction.
It is worth noting that the differences between the various $NN$
potentials are not very relevant, and that second-order corrections
improve remarkably the agreement with experiment.
The latter point confirms the need to go beyond a mean-field
description of ground-state properties of finite 
nuclei\cite{suzuki94,heisenberg99,muther00,fabrocini00}.

\section{Summary and Conclusions} 
In the present work, we have used a new technique to derive a
low-momentum $NN$ potential $V_{low-k}$ from realistic potentials by
integrating out the high-momentum modes. 
With this smooth $V_{low-k}$ we have calculated the bulk properties
of $^{16}$O in the framework of the Goldstone expansion using a
self-consistent Hartree-Fock basis.
The quality of the results, obtained using several modern realistic
$NN$ potentials, is quite good.
We conclude that this new method, that it has also been successfully 
employed in shell-model calculations\cite{bogner02,coraggio02}, may 
contribute to a better understanding of the role of realistic 
$NN$ interactions in nuclear structure.

\section*{Acknowledgments} This work was supported in part by the U.S. DOE 
Grant No. DE-FG02-88ER40388, and the Italian Ministero dell'Istruzione,
dell'Universit\`a e della Ricerca (MIUR).

\end{document}